\documentclass[sigplan,10pt]{acmart}

\acmYear{2021}
\acmISBN{} 
\acmDOI{} 
\startPage{1}

\setcopyright{none}

\usepackage{booktabs}   
\usepackage{subcaption} 
                        
\usepackage[ruled]{algorithm2e}
\usepackage{float}

\begin{document}

\title[Short Title]{Cache Bypassing for Machine Learning Algorithms}         


\author{Asim Ikram}
\affiliation{
  \department{Department of Computer Sciences}              
  \institution{National University of Computer and Emerging Sciences}            
  \city{Islamabad}
  \country{Pakistan}                    
}
\email{asim.ikram1@outlook.com}          

\author{Muhammad Awais Ali}
\affiliation{
  \department{Department of Computer Sciences}             
  \institution{National University of Computer and Emerging Sciences}           
  \city{Islamabad}
  \country{Pakistan}                   
}
\email{m.awaisali.mirza@gmail.com}         

\author{Mirza Omer Beg}
\affiliation{
  \position{Assistant Professor}
  \department{Department of Computer Sciences}             
  \institution{National University of Computer and Emerging Sciences}           
  \city{Islamabad}
  \country{Pakistan}                   
}
\email{omer.beg@nu.edu.pk}         

\begin{abstract}
Graphics Processing Units (GPUs) were once used solely for graphical computation tasks but with the increase in the use of machine learning applications, the use of GPUs to perform general purpose computing has increased in the last few years. GPUs employ a massive amount of threads, that in turn achieve a high amount parallelism, to perform tasks. Though GPUs have a high amount of computation power, they face the problem of cache contention due to the SIMT model that they use. A solution to this problem is called "cache bypassing". This paper presents a predictive model that analyzes the access patterns of various machine learning algorithms and determines whether certain data should be stored in the cache or not. It presents insights on how well each model performs on different datasets and also shows how minimizing the size of each model will affect its performance The performance of most of the models were found to be around 90\% with KNN performing the best but not with the smallest size. We further increase the features by splitting the addresses into chunks of 4 bytes. We observe that this increased the performance of neural network substantially and increased the accuracy to 99.9\% with three neurons.
\end{abstract}



\maketitle

\section{Introduction}
Recent advances in artificial intelligence has led to a renewed attention towards a diverse set of difficult combinatorial problems \cite{beg2013combinatorial,sairarelationship,anwar2020tac,sahar2019towards,zahid2020roman}.
Graphics Processing Units (GPUs) have been used to perform graphics intensive tasks since the earlier days, but in the recent years, thanks to the rise of machine learning, GPUs have been used to perform high performance tasks \cite{khawaja2018domain}. Due to the massive computational power that GPUs offer, they are being used to perform tasks that were once meant to be executed by CPUs. Since GPUs provide more computational power than CPUs, they have been used extensively to train machine learning algorithms\cite{majeed2020emotion,tariq2019accurate,uzair2019weec}.

\begin{figure}[t]
\includegraphics[width=\columnwidth]{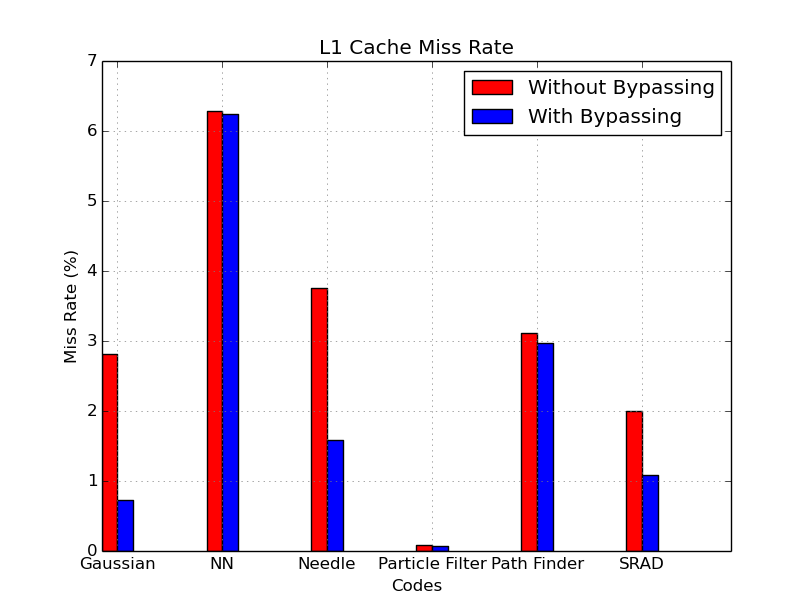}
\caption{Miss rates for L1 cache bypassing on selected Rodinia benchmark programs}
\label{fig:motivation}
\end{figure}

\par Initially, GPUs only had a shared memory called the scratchpad memory installed on board. While scratchpad memory was programmable and enabled rapid fetching of data, it had its own pitfalls. The problem with scratchpad memory was that it performed well on processes that had a uniform access pattern but performed poorly on processes that exhibited irregular access patterns \cite{Huangfu2015}. To handle the problems that scratchpad memory exhibited, vendors started to employ cache memory on GPUs as well.

\par Caches, on GPUs, perform well on tasks that exhibit non-uniform or irregular access patterns. GPU caches do not operate in the same way as CPU caches and hence cannot be optimized in the same way \cite{beg2010graph}. They also exhibit a low level of temporal and spatial locality. While CPU caches have been worked upon in detail, GPU caches still remain an area to be explored \cite{Li2015}.

\par GPUs work by dispatching a large number of threads to streaming multiprocessors (SMs) for processing large datasets in parallel. Since a large number of threads need to be executed, the small size of the cache becomes a bottleneck for performance especially when training deep learning models \cite{naeem2020deep,asad2020deepdetect,javed2020alphalogger}. One problem that arises is called "cache contention". Under a naive approach, the processor would evict data from the cache using the defined eviction policy when the cache reached its maximum capacity. This approach would be ineffective since the data already present in the cache might be more used more frequently than newer data. Since GPUs require a massive amount of threads to be executed in parallel on a shared cache, harmful eviction would likely occur very frequently. One solution to this problem is called "cache bypassing" which involves storing some of the data directly to the L2 cache rather than the L1 cache. 

\par Figure \ref{fig:motivation} illustrates how cache bypassing can improve performance of randomly selected programs running on a GPU. We execute a few programs from the Rodinia benchmark on the Kepler GPU architecture and we modify the PTX data access instructions  to randomly bypass the cache and see improvements on all cases as seen in figure \ref{fig:motivation}.

\par This paper proposes a mechanism that analyzes the access patterns of various machine learning algorithms and uses a model that predicts whether a certain address should be bypassed or not. Furthermore, since the mechanism is intended to be embedded in the hardware, the size of the models has been reduced to a considerable degree to reduce the cost of implementation.  This paper makes the following contributions

\begin{itemize}
\item We propose a cache bypassing mechanism using various machine learning algorithms that learn the access patterns of other machine learning algorithms and takes bypassing decisions.
\item We shrink the size of the model and evaluate its performance. We reduce the size of the model since it is intended to be implemented into the hardware.
\item We further split the addresses into multiple parts and evaluate our models on the modified datasets and find that the performance of the neural network increases substantially.
\end{itemize}

\par The rest of this paper is organized as follows. Section \ref{background} describes the background, section \ref{LR} presents an overview of the related work, section \ref{proposed methodology} describes our methodology, section \ref{experimental setup} lists the experimental setup used to conduct the experiments, section \ref{results} lists the results of the experiments, and section \ref{conclusion} summarizes our work.

\section{Background}\label{background}
\subsection{GPU Architecture}
GPUs consists of multiple parts such Streaming Multiprocessors (SMs). SMs are processors that are designed to handle CUDA requests. Each SM has multiple CUDA cores and a shared L1 data cache \cite{NVIDIACorporation2016}. Each SM is responsible for dispatching warps (blocks of 32 threads) to CUDA cores. 

\subsection{GPGPU}
General Purpose GPU (GPGPU) are GPUs that carry out tasks that are meant to be executed on a CPU. Since GPUs allow massive thread level parallelism, they have been used to perform high intensive computing tasks. Due to the thread handling capabilities of GPUs, lower end GPUs can handle tasks that might require cutting edge CPUs.

\subsection{Cache Bypassing}
GPU caches were introduced to counteract the drawbacks of scratchpad memory. GPU caches perform well on data that exhibits irregular access patterns but while caches have their benefits, they also suffer from some drawbacks. GPUs employ a mechanism called Single Instruction Multiple Threads (SIMT) that involves multiple threads being dispatched for a single process. The small size of the cache becomes a bottleneck when handling such a large number of threads that run in parallel. To increase the performance of the cache in this scenario, cache bypassing can be used. Cache bypassing involves storing only those instructions in the cache that have a high reuse rate, the instructions that do not have a high reuse rate a not stored in the cache and are accessed directly from the memory. This can result in the decrease of cache miss rates.

\section{Related Work}\label{LR}
GPUs have been used in the recent years for high performance computing. To perform this tasks, vendors have started to employ caches for GPUs \cite{Huangfu2015, Xie2015}. The authors claim that due to the nature of GPGPU applications, data data reuse rate is very low and that data can be bypassed to improve cache performance. Yijie Huangfu and Wei Zhang \cite{Huangfu2015} proposed a mechanism that filtered data based on the addresses and their mechanism improved cache performance by 13.8\%. The authors in \cite{Xie2015} classify the data into three types based on locality and use static bypassing for data with high and low level of locality and use dynamic bypassing for data with medium level of locality.

\par The authors in \cite{Liang2015, Dai2016, Xu2015} propose bypassing mechanisms that are compiler based. The authors in \cite{Liang2015} state that only global load instructions are stored in the cache and only those instructions need to be identified and bypassed \cite{Liang2015, Dai2016}. They propose a heuristic based method for a compiler that filters out global load instructions and generates an optimized code. The authors in \cite{Dai2016} propose a model that identifies the optimal number of warps. The authors in \cite{Xu2015} propose 'Hyper Loop Parallelism' to improve the performance of CUDA GPUs. The authors propose a mechanism that identifies whether a loop can be presented in a vector form or not and build a compiler to achieve their goals. \cite{Li2015} also propose a compile time framework to limiting the number of threads that can access a cache. \cite{Park2017} propose a bypass first policy for the last level cache that only stores those addresses in memory that are likely to be re-referenced. \cite{Zhao2017} propose a bypassing scheme that is targeted towards handling un-coalesced loads. The mechanism uses two approaches – One is to bypass data when the number of accesses exceeds a pre-determined threshold. The second approach is to bypass memory accesses when the L1 data cache is stalled. 

\par The authors in \cite{Lee2016, Tian2015} propose bypassing mechanisms that are based on reuses. \cite{Lee2016} propose a mechanism that uses feedback control loops to predict reuse patterns for each instruction. They use a reuse table to keep track of reuses and use data from the table to statistically determine whether to bypass an instruction or not. The result is an almost double speedup. \cite{Tian2015} propose an adaptive cache bypassing mechanism to avoid premature eviction. They use the PC trace to predict bypassing. They predict block that are likely to not be rerefrenced and choose to bypass them which results in a higher hit rate. \cite{Li:2015-1} propose a method that dynamically bypasses instructions and only stores those instructions in the L1 data cache that have a high reuse rate and shore reuse distance. The authors also propose to decouple the L1 data cache to increase the energy efficiency of the cache and to enable the storage of more reuse patterns with a lower overhead.

\section{Proposed Approach}\label{proposed methodology}
The purpose of this paper is to analyze the address patterns of the machine learning algorithms and uses a predictive model to decide whether data should be stored in the cache or not. Since the model is intended to be embedded in the hardware the size of the models should ideally be kept to a minimum.
\par A large number of machine learning models exists and using all of them was not feasible, so we chose a subset of the algorithms available. The algorithms that we ran our test on include \textit{Decision Tree, K Nearest Neighbors (KNN), Logistic Regression, and Neural Network (MLP)}. These machine learning algorithms consist of many parameters to tune and the tuning of these paramaters is an exhaustive approach which is very computationally expensive. To tune these parameters to obtain optimal values, either manually or automatically, takes several days which was can not be considered feasible. Therefore, we optimized these algorithms based on the core component that builds up these algorithms.
For the decision tree, we changed the depth and impurity. We focused on these parameters because the depth controls how many nodes exist in the tree. For increased depths, the nodes in the tree increase along with its overall complexity. When we consider impurity, the computational cost is higher when the impurity is lower. In the case of KNN, we chose to change the value of K since this parameter controls the number of clusters that the model has to construct. For logistic regression, instead of using only one solver, we focused on trying multiple solvers to evaluate the performance of each one and gain insights as to which one performs most optimally on the given datasets. Neural networks have seen a massive amount of use in the recent years. With the introduction of deep learning algorithms as well as the increase in processing power, neural networks have been used in a wide variety of applications such as autonomous driving and IoT \cite{NN1,NN2,NN3,NN4,NN5,NN6}. Since, neural networks have been extremely popular, we were interested in the performance of neural networks with cache bypassing. As our model is intended to be implemented in the hardware, we could not afford to have too many neurons in the network since that would increase the implementation cost. We chose to change the number of neurons in the network while keeping the number of hidden layers to one.

\par The dataset that we used was imbalanced and to balance it we used the SMOTE \cite{SMOTE}. The algorithms we used are given in algorithms ~\ref{alg:DT Depth}, ~\ref{alg:DT Impurity}, ~\ref{alg:KNN}, ~\ref{alg:LR}, ~ \ref{alg:MLP}.

\begin{algorithm}
\SetAlgoLined
 \For{$DepthParam \leftarrow 1$ \KwTo $10$}{
  Initialize Decision Tree($depth=DepthParam$)\;
  evaluate Decision Tree
 }
 \caption{Decision Tree (Depth)}
 \label{alg:DT Depth}
\end{algorithm}

\begin{algorithm}
\SetAlgoLined
 \For{$ImpParam \leftarrow 0$ \KwTo $0.5$}{
  Initialize Decision Tree($impurity=ImpParam$)\;
  evaluate Decision Tree
 }
 \caption{Decision Tree (Impurity)}
  \label{alg:DT Impurity}
\end{algorithm}

\begin{algorithm}
\SetAlgoLined
 \For{$KParam \leftarrow 1$ \KwTo $17$}{
  Initialize KNN($K=KParam$)\;
  evaluate KNN Model
 }
 \caption{KNN}
  \label{alg:KNN}
\end{algorithm}

\begin{algorithm}
\SetAlgoLined
 \For{$SolverParam$ in $[newton-cg, LBFGS, liblinear, sag]$}{
  Initialize Logistic Regression($Solver=SolverParam$)\;
  evaluate Logistic Regression Model
 }
 \caption{Logistic Regression}
  \label{alg:LR}
\end{algorithm}

\begin{algorithm}[h]
 \For{$NeuronParam \leftarrow 1$ \KwTo $20$}{
  Initialize NeuralNetwork($neurons=NeuronParam$)\;
  evaluate Neural Network
 }
 \caption{Neural Network}
  \label{alg:MLP}
\end{algorithm}

\par The dataset that we used consisted of only one feature that was the address. This caused our algorithm to perform not very well, especially for the neural network. To handle this case, we split the address into chunks of 4 bytes and use that data to train the models once more. We present the results on for both versions of the dataset. The algorithm that we used to split our data is given in algorithm \textbf{Splitting algorithm here}

\begin{algorithm}
\SetAlgoLined
	$SplitData \leftarrow []$ \;
	\For{$i$ in $range(sizeOf(Data))$}{
		$record \leftarrow data[i]$ \;
		$tempList \leftarrow []$ \;
		\While{$record \not e 0$}{
			$record, remainder \leftarrow record \% 10$
			$tempList.append(remainder)$
		}
	$SplitData.append(tempList)$
	}
 \caption{Splitting the Addresses}
  \label{alg:Split}
\end{algorithm}

\section{Experimental Setup}\label{experimental setup}
The experiments were conducted on datasets that were generated from tensorflow examples \cite{tensorflow}. The codes that we used are given in table \ref{tab:datasets}. To codes were run on the MNIST dataset \cite{mnist}. We ran several iterations on each of the examples and generated a dataset that contained approximately 1,000,000 records.

\begin{table}[h]
\caption{Codes Used for Generating Datasets}
\label{tab:datasets}
\begin{center}
\begin{tabular}{@{}l@{}}
\toprule
\textbf{Code Used} \\
\midrule
Nearest Neighbors \\
Logistic Regression \\
Random Forest \\
Recurrent Neural Network \\
\bottomrule
\end{tabular}
\end{center}
\end{table}

\par The models that we used for training are given in section \ref{proposed methodology}. The models were trained on a machine with an Intel i7 3630QM processor (2.4Ghz) 8GB RAM, and an Nvidia Geforce GT 630M \cite{gt630m}. The specifications for the GPU are given in table \ref{tab:gpu specs}. The models were implemented using using Python 3.6.4. As the model(s) would be implemented at the hardware level, we reduced the size of the models to reduce implementation costs while maintaining an acceptable level of accuracy.

\begin{table}[h]
\caption{Geforce GT 630M Specifications}
\label{tab:gpu specs}
\begin{center}
\begin{tabular}{ll}
\toprule
\textbf{Item} 				& \textbf{Value} \\
\midrule
CUDA Cores  				& 96 \\
Graphics Clock				& 800MHz \\
Memory Interface Width		& 128 bits \\
Architecture				& Kepler \\
Memory Bandwidth			& 32 GB/s \\
Memory Size					& 2 GB \\
\bottomrule
\end{tabular}
\end{center}
\end{table}

\section{Results}\label{results}
We tested our models by using data generated from the tensorflow examples mentioned earlier. We measured the accuracy of each model with respect to the model's size. The results for each of the datasets on different models are given in the subsequent sections.

\subsection{Case 1: One Feature}

\subsubsection{Logistic Regression Dataset}

\begin{figure*}[ht]
\includegraphics[width=\textwidth]{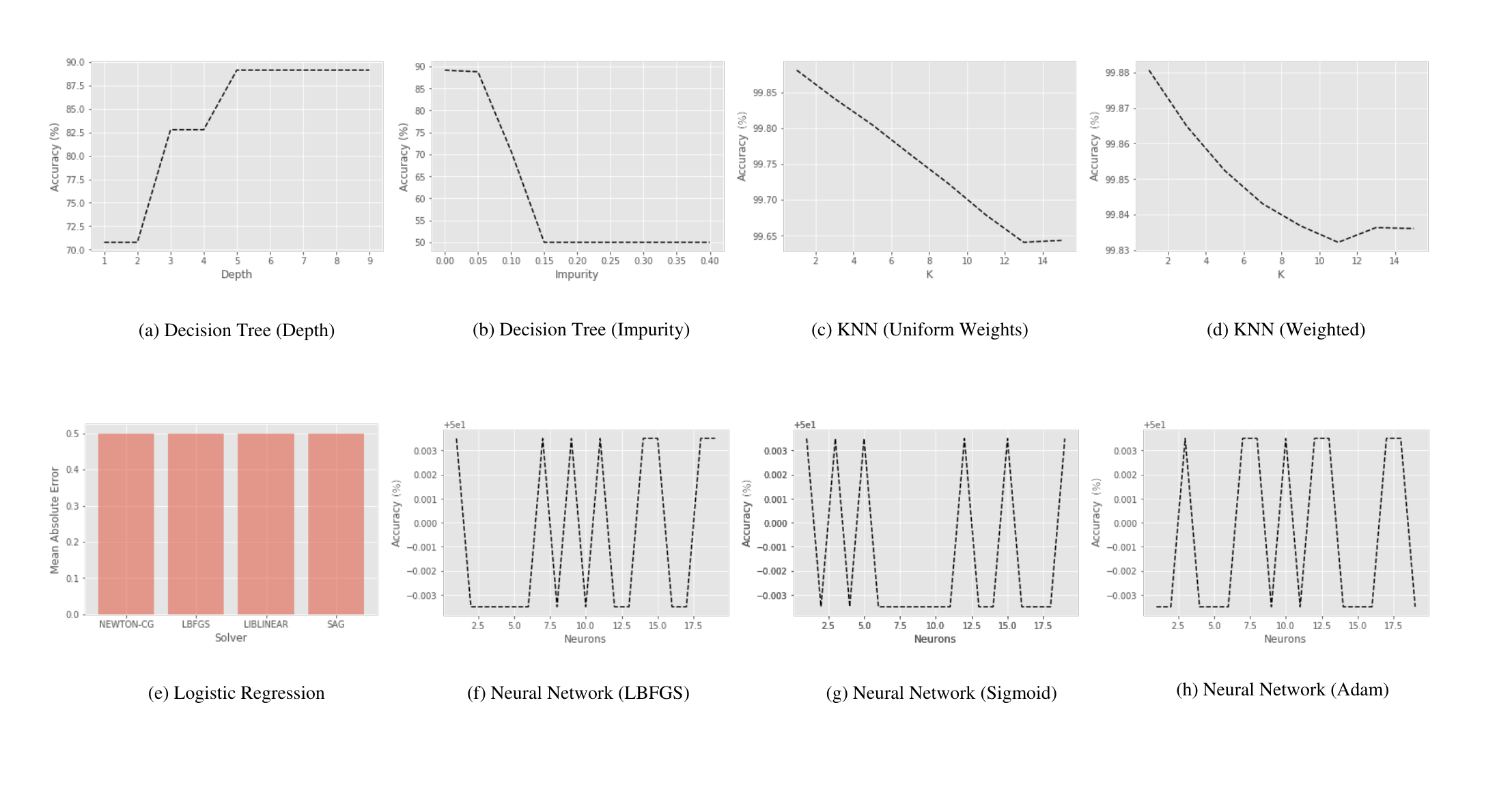}
\hrule
\caption{Results on Logistic Regression Dataset}
\label{fig:LR Dataset}
\end{figure*}

\par Figure \ref{fig:LR Dataset} shows the results obtained from the machine learning models on the logistic regression dataset. For the decision tree, when the impurity was kept constant and the depth of the tree was changed it was observed that the accuracy varied when the depth was changed from 1 to 5 but from depth 5 onwards the accuracy remained the same. When the depth was kept constant and the impurity was changed, it was observed that the accuracy was high when the impurity was lower than 0.05 but dropped drastically when the impurity was increased beyond this point. When the impurity was set to greater than or equal to 0.15, the accuracy remained the same. For KNN, the accuracy decreased more sharply in the case of uniform weights.
All of the versions of the neural networks had an accuracy of approximately 50\% and exhibited an irregular behavior when the number of neurons in the network was changed. A notable point was that the sigmoid and lbfgs solvers gave the highest accuracy with one neuron while the Adam solver showed a converse behavior.
For logistic regression, all the solvers gave the same error. This implies that the solver most suited for the situation could be used without affecting the performance of the model.

\subsubsection{Nearest Neighbors Dataset}

\begin{figure*}[h!]
\includegraphics[width=\textwidth]{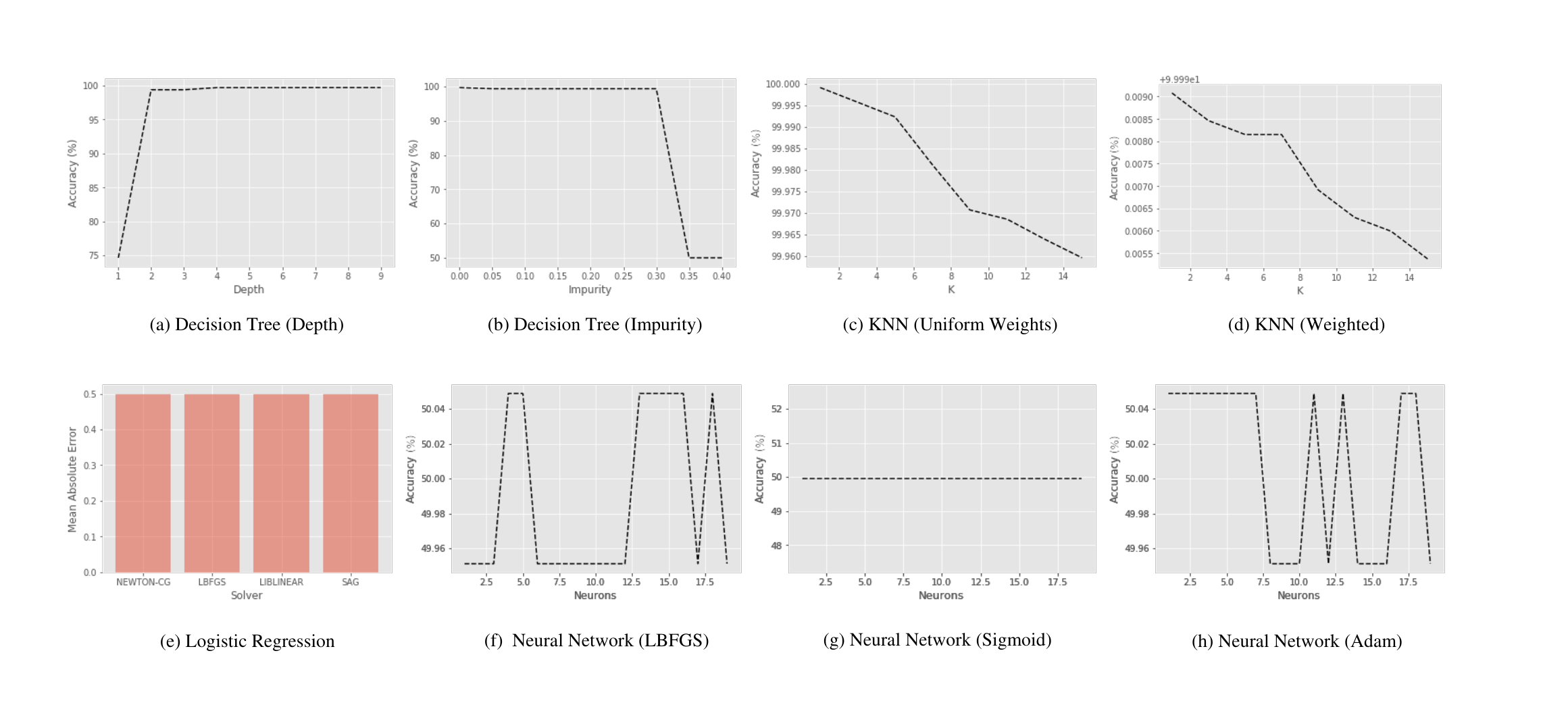}
\hrule
\caption{Results on the Nearest Neighbors Dataset}
\label{fig:NN Dataset}
\end{figure*}

\par Figure \ref{fig:NN Dataset} shows the results obtained on the nearest neighbors dataset using the machine learning  algorithms. In the case of the decision tree, the accuracy increased suddenly when the depth was increased from 1 to 2 and increased very slightly when the depth was increased from 3 to 4. The accuracy remained constant at depths greater then or equal to 4 (impurity was constant). When the depth was kept constant and the impurity was changed, the accuracy was very high when the impurity was below 0.30 but decreased abruptly when the impurity increased beyond this point. This implies that the nearest neighbors dataset is relatively resilient to impure splitting.
The accuracy obtained on the same dataset using both versions of KNN was very high. It can be seen from the graphs that accuracy decreased more consistently with increasing values of K when using uniform weights.
The mean absolute error obtained using logistic regression with different solvers on the nearest neighbors dataset was the same on all the solvers.
For the neural network, the sigmoid solver exhibited a constant accuracy when the number of neurons in the network was changed. The LBFGS and Adam solvers displayed a completely random pattern no matter how many neurons existed in the network. In this case, the Adam solver achieved the highest accuracy with one neuron while LBFGS exhibited the lowest accuracy with one neuron.

\subsubsection{Random Forest Dataset}

\begin{figure*}[h!]
\centering
\includegraphics[width=\textwidth]{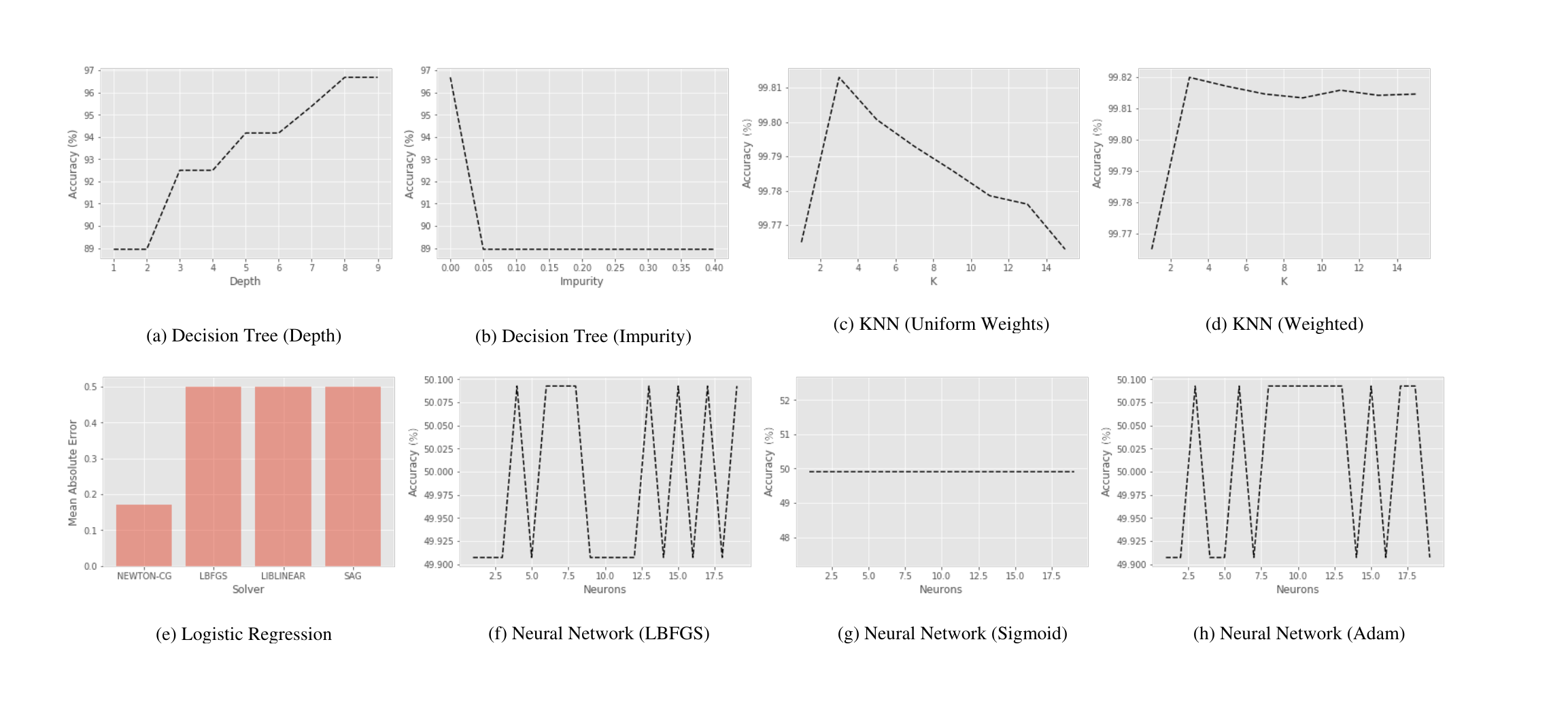}
\hrule
\caption{Results on the Random Forest Dataset}
\label{fig:RF Dataset}
\end{figure*}

\par Figure \ref{fig:RF Dataset} shows the results obtained on the random forest dataset using the machine learning algorithms. I In the case of the decision tree, when the depth was changed and impurity was kept constant, the accuracy increased at approximately equal intervals. When the depth was kept constant and the impurity was changed the accuracy dropped abruptly when the impurity was increased from 0 and stayed constant at values equal to or above 0.05. This implies that accuracy decreases substantially when the splitting of nodes is not pure.
For both of KNN, the accuracy is the highest when K is 3 but the accuracy decreases sharply in the case of uniform weights and remains almost constant in the case of weighted KNN.
When using logistic regression, the error on newton-cg was considerably lower then the other three solvers. The error on LBFGS, liblinear and sag was the same.
In the case of neural networks, the sigmoid solver showed a behavior separate from the rest and exhibited a constant accuracy no matter how many neurons existed in the network. The LBFGS and adam solvers showed an irregular behavior when the number of neurons was changed.

\subsubsection{Recurrent Neural Network Dataset}

\begin{figure*}[h!]
\centering
\includegraphics[width=\textwidth]{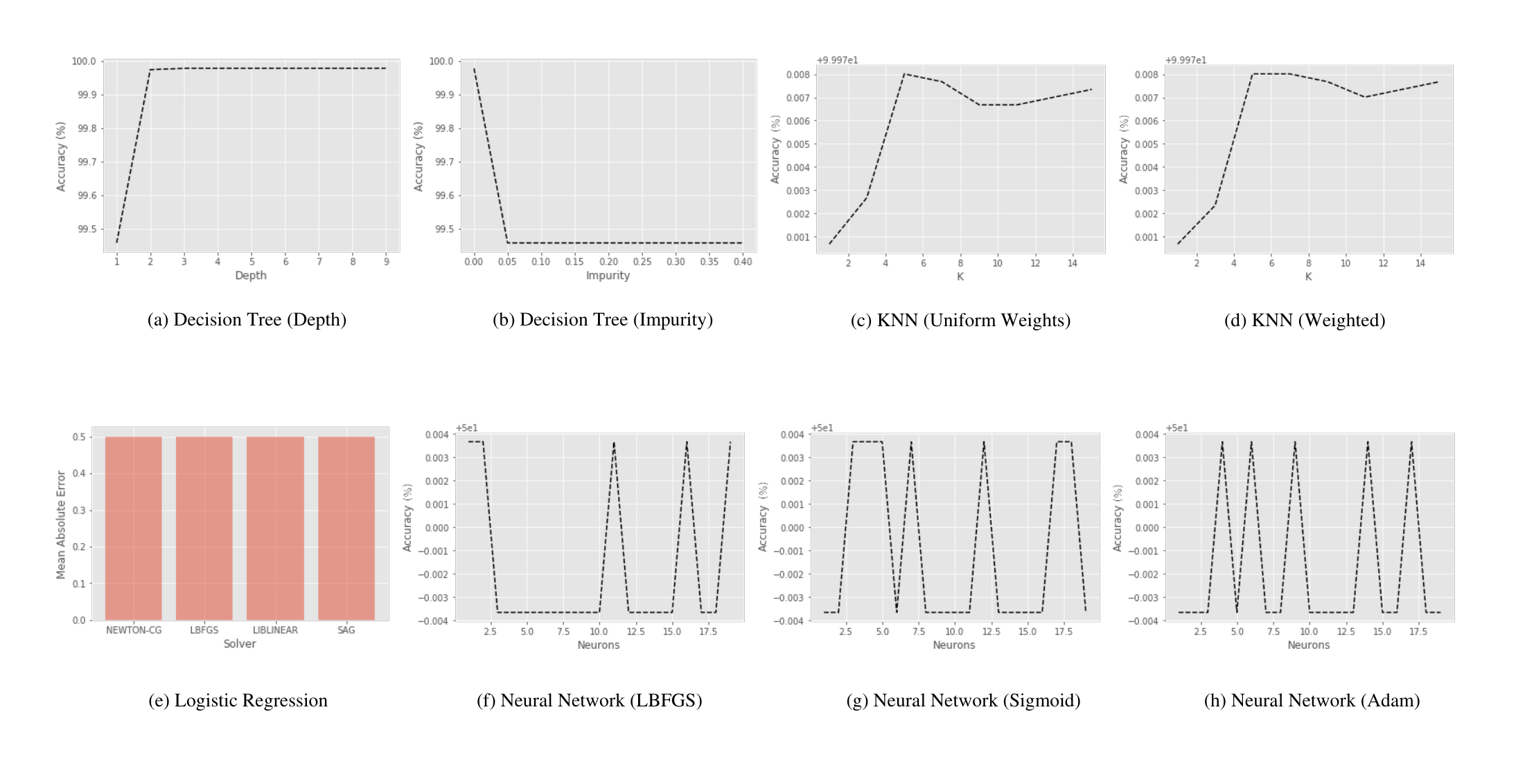}
\hrule
\caption{Results on the RNN Dataset}
\label{fig:RNN Dataset}
\end{figure*}

\par Figure \ref{fig:RNN Dataset} shows the results obtained achieved on the RNN dataset using a the models. For the decision tree, the accuracy increased substantially when the depth was increased from 1 to 2 but increased by a very minor amount when the depth was increased from 2 to 3. The accuracy remained constant at depth greater than or equal to 3 (impurity was kept constant). When the depth was kept constant and the impurity was increased, the accuracy decreased rapidly when the impurity was increased from 0 meaning that a slightly impure split can cause a rapid decrease in accuracy on this dataset.
For both versions of KNN, the highest accuracy was achieved when the value of K was 5. For values of K greater than 5, the behavior of both versions was almost the same.
The error on all the solvers remained the same when using logistic regression. This implies that for the RNN dataset, any of the solvers could be applied (most preferably the best one for the situation) without affecting the error.
For the neural network, all of the solvers showed a random behavior when the number of neurons in the network was changed. The LBFGS solver achieved the highest accuracy when one neuron existed in the network. The sigmoid and Adam solvers exhibited the opposite behavior under the same conditions.

\subsection{Case 2: Multiple Features}
At first we tested our model on the original dataset that consisted of only one feature. Since using one feature was causing a linear mapping, we decided to split the data into parts to make a more generic model and tested our models on the modified dataset as well. The results of the modification are described in the subsequent sections.

\subsubsection{Logistic Regression Dataset}

\begin{figure*}[h!]
\centering
\includegraphics[width=\textwidth]{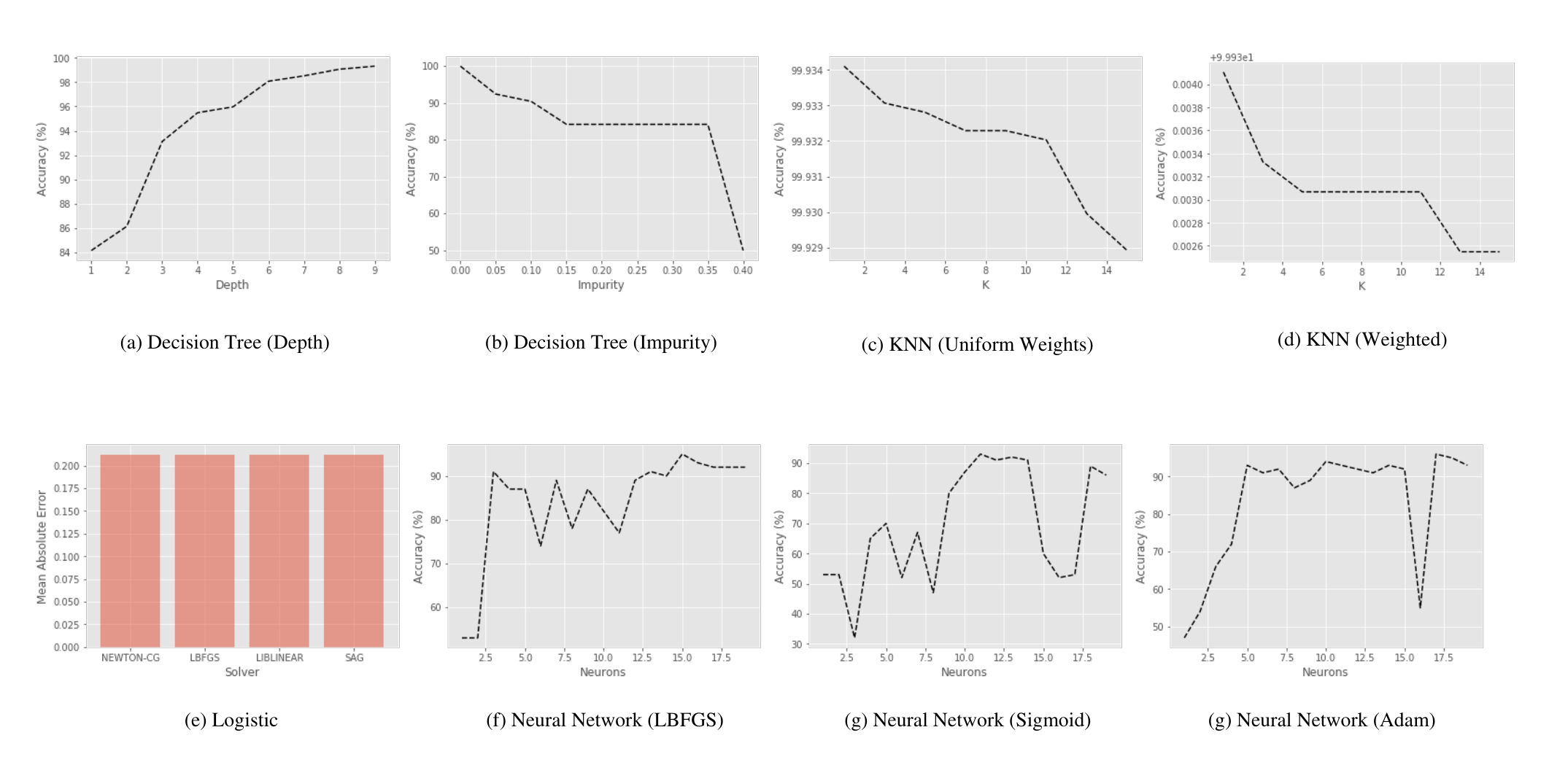}
\hrule
\caption{Results on the Logistic Regression Dataset (Case 2)}
\label{fig:LR Dataset 2}
\end{figure*}

Figure \ref{fig:LR Dataset 2} shows the results obtained from the machine learning models on the logistic regression dataset having multiple features. For the decision tree, when the impurity was kept constant and the depth of the tree was changed it was observed that the accuracy varied when the depth was changed from 1 to 9 and kept increasing. When the depth was kept constant and the impurity was changed, it was observed that the accuracy was high when the impurity was lower than 0.05 but dropped drastically when the impurity was increased beyond this point. When the impurity was set to greater than or equal to 0.15, the accuracy remained the same. For KNN, the accuracy decreased more sharply in the case of uniform weights. All of the versions of the neural networks had an accuracy of approximately 50\% and exhibited an irregular behavior when the number of neurons in the network was changed. A notable point was that the sigmoid and lbfgs solvers gave the highest accuracy with two neurons while the highest accuracy was achieved with five neurons using Adam solver. For logistic regression, all the solvers gave the same error. This implies that the solver most suited for the situation could be used without affecting the performance of the model.

\subsubsection{Nearest Neighbors Dataset}

\begin{figure*}[h!]
\centering
\includegraphics[width=\textwidth]{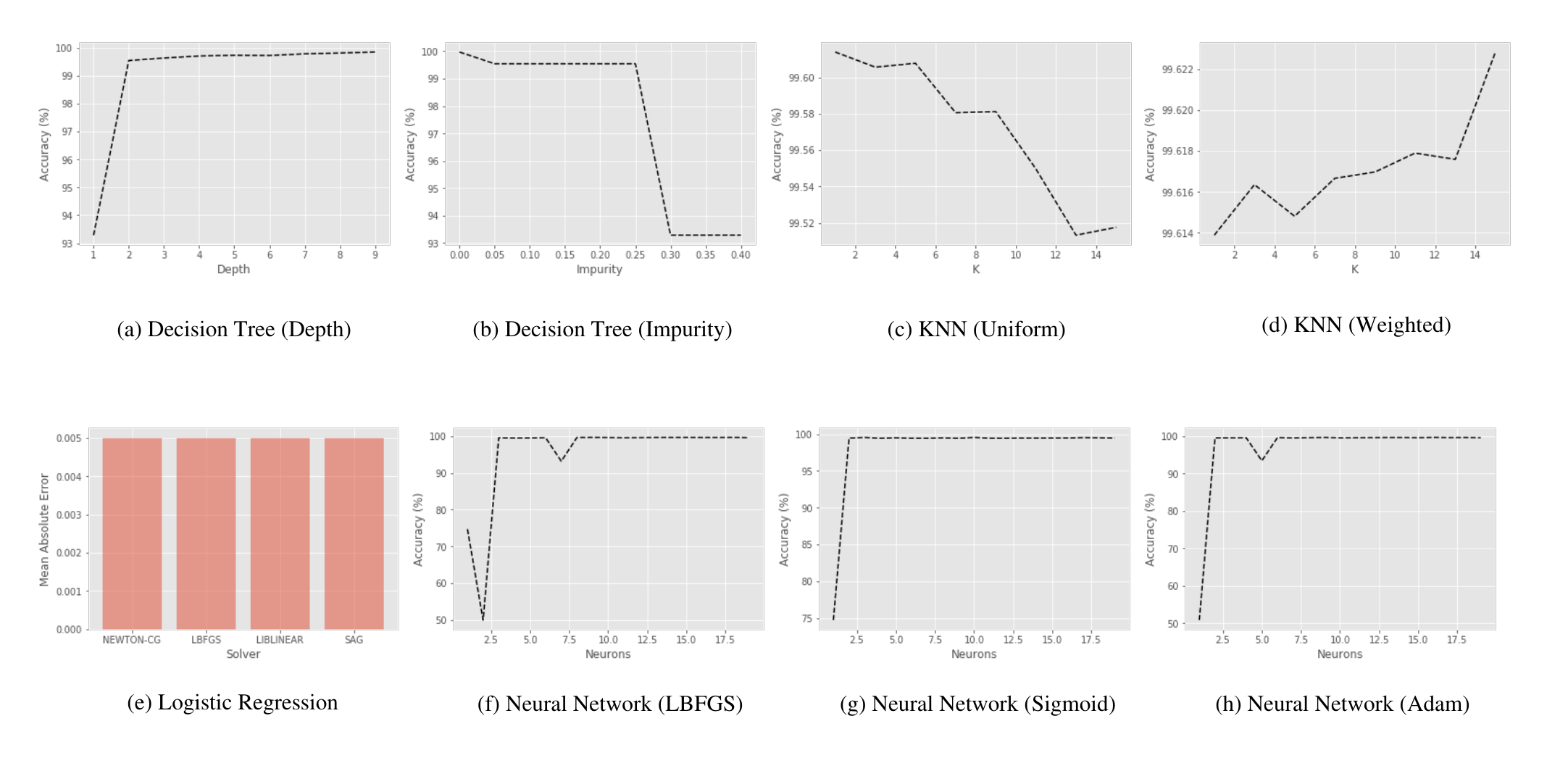}
\hrule
\caption{Results on the Nearest Neighbors Dataset (Case 2)}
\label{fig:KNN Dataset 2}
\end{figure*}

Figure \ref{fig:KNN Dataset 2} shows the results obtained on the nearest neighbors dataset with multiple features .In the case of the decision tree, the accuracy increased suddenly when the depth was increased from 1 to 2 and increased very slightly when the depth was increased from 3 to 4. The accuracy remained constant at depths greater then or equal to 4 (impurity was constant). When the depth was kept constant and the impurity was changed, the accuracy was very high when the impurity was below 0.0.05 but decreased slightly when the impurity increased beyond this point and then abruptly when it was increased beyond 0.25. This implies that the nearest neighbors dataset is relatively resilient to impure splitting. The accuracy obtained on the same dataset using both versions of KNN was very high. It can be seen from the graphs that accuracy decreased more consistently with increasing values of K when using uniform weights and it increased gradually with weighted KNN as the number of K was increased. The
mean absolute error obtained using logistic regression with different solvers on the nearest neighbors dataset was the same on all the solvers. For the neural network, the sigmoid solver exhibited a constant accuracy when the number of neurons in the network was changed. The LBFGS and Adam solvers displayed almost similar pattern. In LBFGS the accuracy was initially 75\% with one neuron and then there is a sudden decrease when the number of neurons were increased to 2 but after that showed the same behavior as Adam, it increased gradually as the number of neurons were increased.

\subsubsection{Random Forest Dataset}

\begin{figure*}[h!]
\centering
\includegraphics[width=\textwidth]{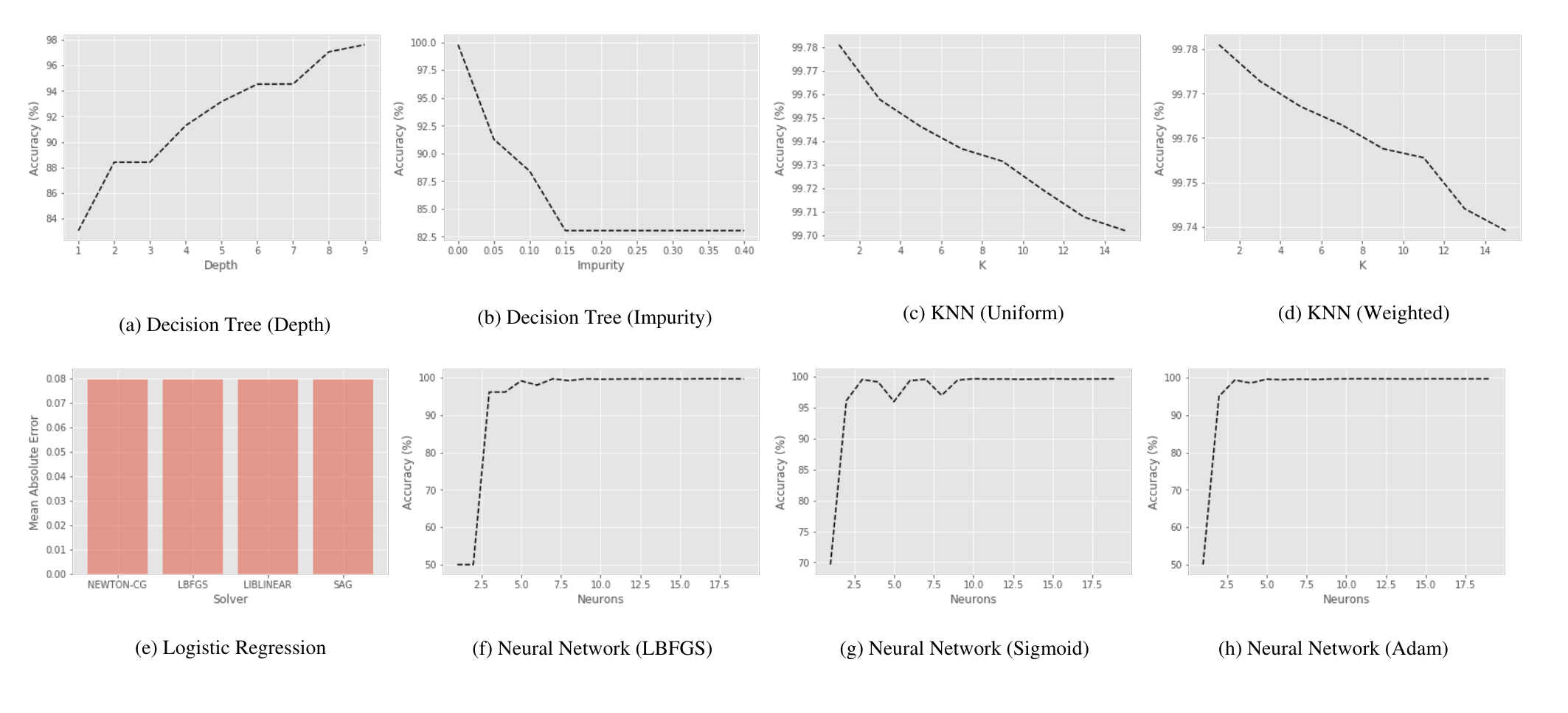}
\hrule
\caption{Results on the Random Forest Dataset (Case 2)}
\label{fig:RF Dataset 2}
\end{figure*}

Figure \ref{fig:RF Dataset 2} shows the results obtained on the random forest dataset with multiple features using the machine learning algorithms. In the case of the decision tree, when the depth was changed and impurity was kept constant, the accuracy increased gradually. When the depth was kept constant and the impurity was changed the accuracy dropped abruptly when the impurity was increased from 0 and stayed constant at values equal to or above 0.15. This implies that accuracy decreases substantially when the splitting of nodes is not pure. For the KNN, the accuracy is the highest when K is 1 but the accuracy decreases drastically in both cases when K was increased. When using logistic regression, the error on all three solvers remain the same.In the case of neural networks, all solvers exhibit the same behavior as the number of neurons increased. Neural network accuracy drastically increases and then kept constant as the neurons were further increased.

\begin{figure*}[th!]
\centering
\includegraphics[width=\textwidth]{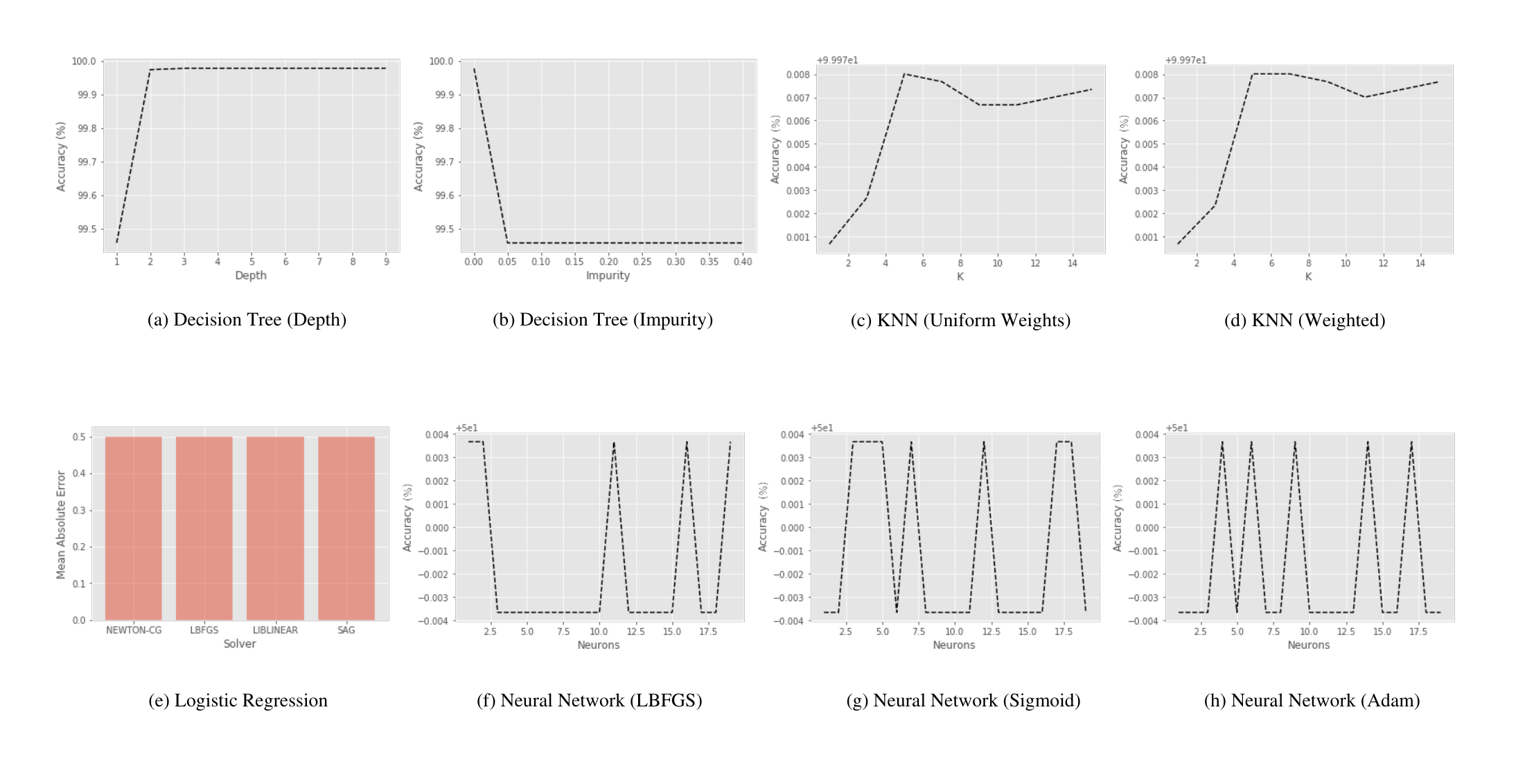}
\hrule
\caption{Results on the RNN Dataset}
\label{fig:RNN Dataset 2}
\end{figure*}
\subsubsection{Recurrent Neural Network Dataset}

Figure ~\ref{fig:RNN Dataset 2} shows the results achieved on the RNN dataset with multiple features. For the decision tree, the accuracy increased substantially when the depth was increased from 1 to 2 but then remained constant at depth greater than 2. When the depth was kept constant and the impurity was increased, the accuracy decreased rapidly when the impurity was increased from 0 meaning that a slightly impure split can cause a rapid decrease in accuracy on this dataset. For both versions of KNN, the highest accuracy was achieved when the value of K was 5. For values of K greater than 5, the behavior of both versions was almost the same. The error on all the solvers remained the same when using logistic regression except for the LIBLINEAR, which shows a slight decrease in error than the others. This implies that for the RNN dataset, LIBLINEAR solvers could be applied (most preferably the best one for the situation) without affecting the error. For the neural network, LBFGS shows the highest accuracy when 3 neurons were used and then it kept constant as the number of neurons were increased. The sigmoid and Adam solvers exhibited an irregular behavior when the number of neurons in the network was changed.

\section{Conclusion}\label{conclusion}
GPUs have been used in the recent years to perform general purpose computing tasks. For this purpose, caches have been added to GPUs but they perform poorly due to the massive number of threads accessing the cache. This paper analyzes different machine learning algorithms and presents insights on whether to bypass or cache addresses using various machine learning algorithms. It also presents how reducing the size of the machine learning algorithms will affect their performance.


\bibliography{bypassing}



\end{document}